\documentclass[12pt,twocolumn,letterpaper]{article}

\usepackage{graphicx}
\usepackage{amsmath}
\usepackage{amssymb}
\usepackage{booktabs}
\usepackage{wacv}
\usepackage{times}
\usepackage{epsfig}
\usepackage{graphicx}
\usepackage{amsmath}
\usepackage{amssymb}
\usepackage{booktabs}
\usepackage{overpic}%
\usepackage{caption}
\usepackage{algorithm}%
\usepackage{algorithmicx}%
\usepackage{algpseudocode}
\usepackage{enumitem}

\usepackage[pagebackref,breaklinks,colorlinks]{hyperref}

\usepackage[capitalize]{cleveref}
\crefname{section}{Sec.}{Secs.}
\Crefname{section}{Section}{Sections}
\Crefname{table}{Table}{Tables}
\crefname{table}{Tab.}{Tabs.}

\begin{document}

\title{Digital Kitchen Remodeling:\\ Editing and Relighting Intricate Indoor Scenes\\ from a Single Panorama}

\author{author\\
institution\\
{\tt\small email}
\and
author\\
institution\\
{\tt\small email}
\and
author\\
institution\\
{\tt\small email}
}

\author{Guanzhou Ji\\
Carnegie Mellon University\\
{\tt\small gji@andrew.cmu.edu}
\and
Azadeh O. Sawyer\\
Carnegie Mellon University\\
{\tt\small asawyer@andrew.cmu.edu}
\and
Srinivasa G. Narasimhan\\
Carnegie Mellon University\\
{\tt\small srinivas@andrew.cmu.edu}
}

\twocolumn[{
    \renewcommand\twocolumn[1][]{#1}%
    \maketitle
    \begin{center}
      \centering
      \begin{overpic}[width=\textwidth]{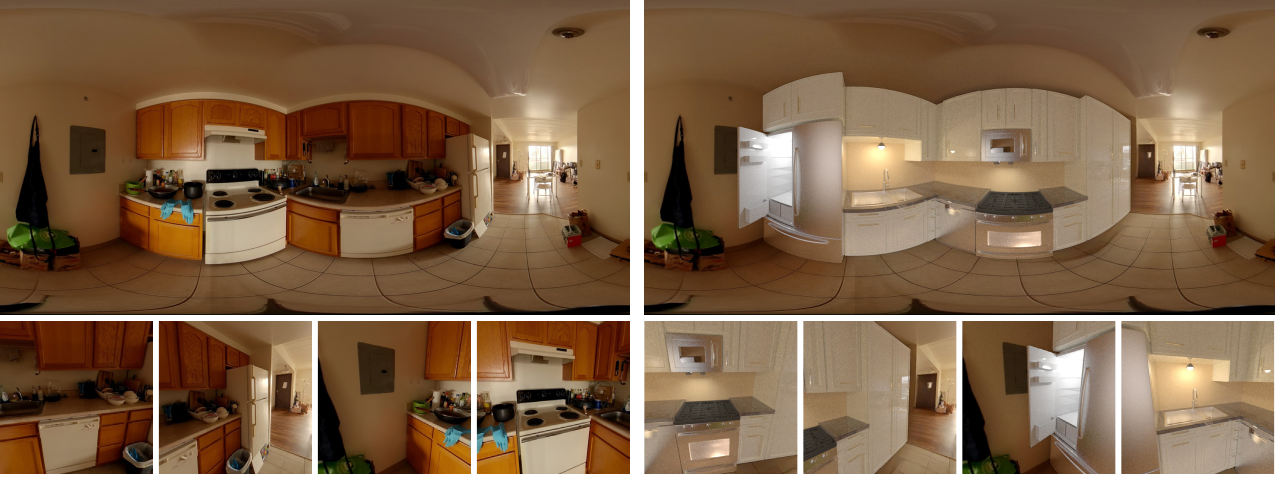}
        \put(18,38){\small \color{black}{Captured Scene}}
        \put(71,38){\small \color{black}{Edited Scene}}
      \end{overpic}
      \captionof{figure}{Kitchen Remodeling Application: (Left) A captured indoor panorama showcases an intricate existing scene. (Right) The kitchen area is virtually remodeled with new walls, cabinets and appliances, and is remodeled within the existing outdoor context with additional electrical lighting.}
      \label{fig_teaser}
    \end{center}
}]

\begin{abstract}
We present a novel virtual staging application for kitchen remodeling from a single panorama. To ensure the realism of the virtual rendered scene, we capture real-world High Dynamic Range (HDR) panoramas and recover the absolute scene radiance for high-quality scene relighting. Our application pipeline consists of three key components: (1) HDR photography for capturing paired indoor and outdoor panoramas, (2) automatic kitchen layout generation with new kitchen components, and (3) an editable rendering pipeline that flexibly edits scene materials and relights the new virtual scene with global illumination. Additionally, we contribute a novel Pano-Pano HDR dataset with 141 paired indoor and outdoor panoramas and present a low-cost photometric calibration method for panoramic HDR photography. The project webpage for this work can be seen \href{https://j-review.github.io/kitchen/}{here}.
\end{abstract}

\section{Introduction}
\label{sec:intro}

Virtual staging techniques digitally showcase indoor scenes. With U.S. home sales projected to reach 5.6 million in 2025~\cite{statista2024homes}, these techniques provide a convenient way for the public to evaluate the property by changing the viewpoint, lighting, furniture, paint, flooring and other attributes. The recent popularity of panoramic cameras has driven the growth of virtual home tours~\cite{zillow2020home}. A single panorama has the advantage of providing a 360$^{\circ}$ view for immersive visualization.

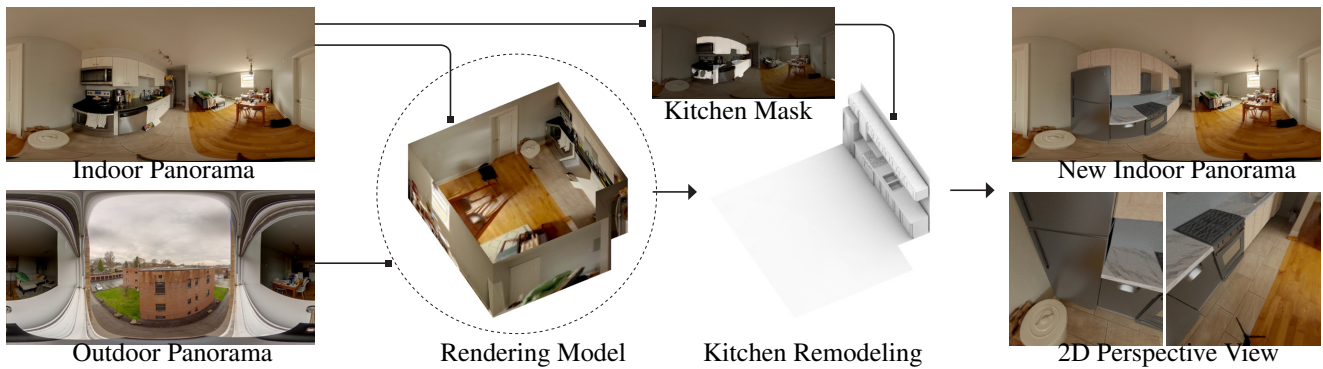
\begin{figure*}
  \centering
  \begin{overpic}[width=\textwidth]{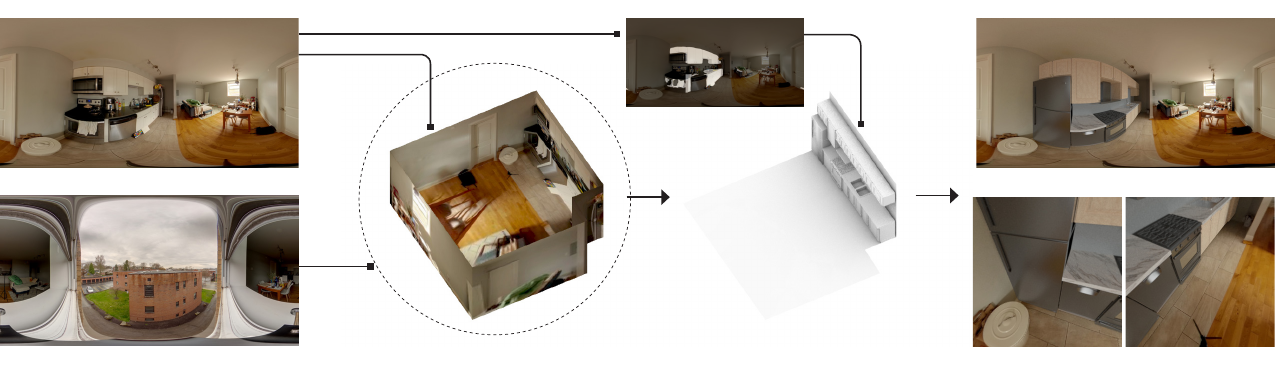}
    \put(5,15){\footnotesize \color{black}{Indoor Panorama}}
    \put(5,1){\footnotesize \color{black}{Outdoor Panorama}}
    \put(33,1){\footnotesize \color{black}{Rendering Model}}
    \put(50,19.5){\footnotesize \color{black}{Kitchen Mask}}
    \put(53,1){\footnotesize \color{black}{Kitchen Remodeling}}
    \put(80,15){\footnotesize \color{black}{New Indoor Panorama}}
    \put(80,1){\footnotesize \color{black}{2D Perspective View}}
  \end{overpic}
  \caption{Overview of Our Approach: We present a comprehensive pipeline from data collection to scene editing and relighting. First, a single Ricoh Theta Z1 camera captures paired indoor and outdoor panoramas. Second, the indoor panorama is used to estimate the 3D room layout, and the outdoor panorama provides a 360$^{\circ}$ environment map. Third, the kitchen area is segmented from the indoor panorama, and new virtual kitchen components are positioned along the kitchen wall. Finally, the new kitchen space is rendered under global illumination, and the single panorama can be converted into 2D perspective views.}
  \label{fig_workflow}
\end{figure*}

Virtual home staging presents several technical challenges. First, the variations in indoor and outdoor lighting make scene relighting difficult. Under natural illumination, outdoor spatially-varying light enters through window openings, illuminates the scene, and shapes the indoor appearance. Second, virtual staging for intricate spaces, such as kitchens, is particularly challenging. Kitchen spaces feature multiple components, such as cabinets and appliances, arranged in a small area according to interior layout rules. The range of material properties and localized lighting in different task areas adds to the complexity of scene relighting. The goal of this work is to edit and relight intricate spaces to achieve high-quality virtual staging.

Existing studies on 2D perspective images primarily focus on estimating global light and relighting objects within the scene. Given a single perspective, previous research has focused on estimating 360$^{\circ}$ HDR environment maps directly from Low Dynamic Range (LDR) images to relight and insert new virtual objects ~\cite{gardner2019deep,gkitsas2020deep,legendre2019deeplight,gardner2017learning,dastjerdi2023everlight}. Given multi-view perspective, light estimation provides flexibility for novel view changes~\cite{li2023multi,philip2021free}. Using indoor HDR environment maps solely as light sources to relight virtual~\cite{wu2023factorized} and real-world scenes~\cite{li2022physically,gardner2017learning,dastjerdi2023everlight}, these image-based rendering methods cannot recover appearances due to complex outdoor lighting. Thus, scene relighting tasks have primarily focused on inserting small objects, such as a metal sphere or a point light source. 

Recent studies have begun to estimate the effect of outdoor lighting using a single panorama. For empty scenes, Zhi et al.~\cite{zhi2022semantically} focus on decomposing light effects in the indoor scene due to the sun and sky and editing the light directions. This study shows simple floor layouts with the scene under direct illumination. For furnished scenes, some studies focus on furniture removal tasks in virtual panoramas~\cite{gkitsas2021panodr,gkitsas2021towards} and real-world panoramas~\cite{matterport2024defurnishing}. To ensure realistic virtual staging, Ji et al.~\cite{ji2023virtual,ji2024virtual} use paired indoor panoramas and outdoor hemispherical images to virtually render the existing scenes with simple floor layouts, materials, and illumination conditions. However, this approach requires two distinct cameras and photometric calibration with absolute light measurement for both cameras, leading to higher labor and financial costs.

This paper presents an application for editing intricate kitchen spaces from a single panorama (Fig.~\ref{fig_teaser}). We use a single camera device (Ricoh Theta Z1) to capture paired indoor and outdoor panoramas with the HDR technique. Given that actual scene radiance depends on the individual scene and necessitates on-site measurements, we introduce a new low-cost approach that calibrates each HDR image to accurate brightness, making our HDR images different from the existing indoor panorama datasets~\cite{xiao2012recognizing,zhang2014panocontext,cruz2021zillow,ji2024virtual,ji2023virtual}. The calibrated outdoor HDR panorama serves as an environment map. From a single indoor panorama, we construct a complete rendering model using 3D scene geometry, material properties, and outdoor illumination. To automate the kitchen remodeling process, we contribute a kitchen component dataset that allows us to edit material properties for editable light estimation.

Overall, we introduce a virtual staging application, from data collection to scene relighting, to edit and relight intricate indoor scenes (Fig.~\ref{fig_workflow}). Specifically, this paper makes the following technical contributions:

\begin{itemize}[noitemsep]
  \item An inverse rendering pipeline using a single camera (Ricoh Theta Z1) to capture paired indoor and outdoor HDR panoramas to relight intricate indoor scenes.
  \item A low-cost photometric calibration method that linearly scales each HDR panorama to display the actual scene radiance. 
  \item A calibrated HDR dataset comprising 141 paired indoor-outdoor panoramas. Each scene is labeled with photometric measurements (indoor illuminance, outdoor illuminance, and indoor luminance) and room orientation.
  \item A kitchen remodeling application that automatically transforms existing kitchen spaces and allows for the customization of material properties for new virtual objects, with a new 3D dataset of virtual kitchen components.
\end{itemize}

\section{Related Work}
\subsection{Indoor Layout Estimation}
Indoor layout estimation from a single panorama is a common task in computer vision. Building on the Manhattan world assumption~\cite{coughlan1999manhattan}, previous studies have explored both cubic~\cite{zhao20223d,cheng2018cube,zioulis2021single} and non-cuboid layouts~\cite{rao2021omnilayout,wang2021led2}. However, due to occlusions inherent in a single viewpoint, a single panorama may capture only a partial view of the space, necessitating the use of two panoramas to estimate the full layout~\cite{wang2022psmnet}. Additionally, there has been research on floor estimation in environments with multiple rooms~\cite{solarte2022360}. While these works primarily focus on indoor spaces with regular geometry, layout estimation and scene segmentation for more complex environments, such as kitchen spaces, remain largely unexplored.

\subsection{Indoor Scene Relighting}
Inserting virtual objects into images is a common task in computer vision and graphics. Previous studies primarily use learning-based approaches to directly estimate the 360$^{\circ}$ environment map~\cite{gardner2017learning,gkitsas2020deep,gardner2019deep} from 2D perspective images. In addition to lighting, surface planes, such as walls, floors, and ceilings, are estimated to insert shadows along with new objects~\cite{li2022physically,li2020inverse,yeh2022photoscene}. User inputs provide additional information to define scene geometry and light direction~\cite{karsch2011rendering,zhang2016emptying}. However, due to the incomplete scene layout, the results often lack the inter-reflection between new objects and the existing scene. A recent study by Li et al.~\cite{li2021lighting} estimates spatially-varying lighting, reflectance, and 3D geometry from a single panorama. However, it is challenging to estimate the complete light transport from outdoor to indoor space using just a single indoor panorama.

\section{HDR Dataset and Low-cost Photometric Calibration}
We present a data collection approach to capture paired indoor and outdoor panoramas, with the outdoor HDR photograph taken immediately after the indoor photograph. A single Ricoh Theta Z1 camera was used to capture panoramic HDR photographs, with multiple exposures controlled by a smartphone and the multi-bracketing function. The camera settings follow previous studies~\cite{ji2023virtual,ji2024virtual}.

\begin{figure}[!b]
  \begin{overpic}[width=\linewidth]{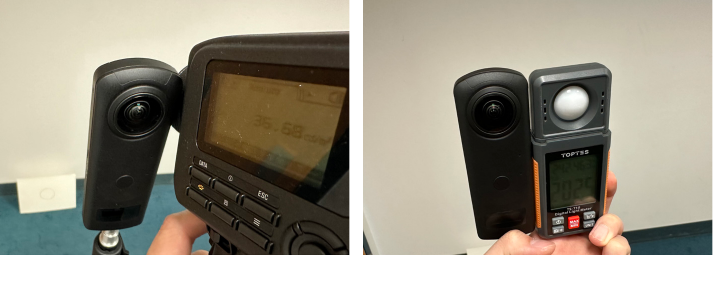}
    \put(22,0){\small \color{black}{(a)}}
    \put(72,0){\small \color{black}{(b)}}
  \end{overpic}
   \caption{Two Photometric Measurements: (a) A Konica Minolta LS-160 luminance meter (over \$5{,}000\ US dollars) is placed by the Ricoh Theta Z1 camera lens to measure the target luminance value on the whiteboard, as used in previous studies~\cite{ji2023virtual,ji2024virtual}. (b) Our low-cost approach: A TS-710 light meter (around \$30\ US dollars) is placed by the Ricoh Theta Z1 camera lens to measure the illuminance coming from 180$^{\circ}$ hemispherical directions.}
\label{fig_hdr_cali}
\end{figure}


Each HDR photograph requires radiometric self-calibration~\cite{mitsunaga1999radiometric} to accurately display scene radiance. Using the standard photometric measurement (Fig.~\ref{fig_hdr_cali} (a)), a Konica Minolta LS-160 luminance meter is placed next to the camera to measure the target absolute luminance ($cd/m^{2}$) at the center of a matte whiteboard. The measured luminance value is then compared with the luminance value displayed in the captured HDR photograph to compute the scaling factor for per-pixel calibration~\cite{kumaragurubaran2013hdrscope}. 

\begin{figure}
  \begin{overpic}[width=\linewidth]{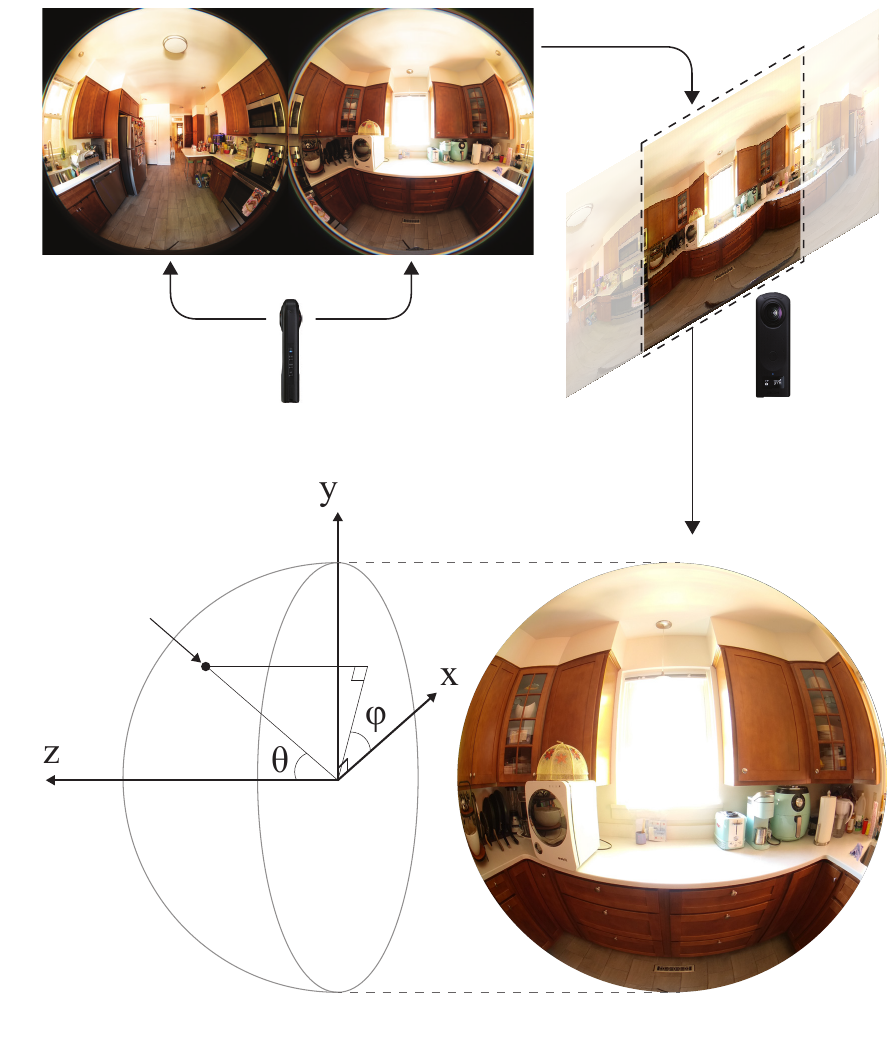}
    \put(17,67){\footnotesize \color{black}{lens 1}}
    \put(32,67){\footnotesize \color{black}{lens 2}}
    \put(17,58){\footnotesize \color{black}{Ricoh Theta Z1}}
    \put(0,85){\footnotesize \color{black}{(a)}}
    \put(0,35){\footnotesize \color{black}{(b)}}
    \put(10,1){\footnotesize \color{black}{Hemispherical Model}}
    \put(43,1){\footnotesize \color{black}{Orthographic Fisheye Image}}
  \end{overpic}
  \caption{Using Illuminance Measurement to Calibrate HDR Panorama. (a) The content captured by the front lens is cropped and geometrically transformed into a 180$^{\circ}$ orthographic fisheye image. (b) Scene illuminance measured from the 180$^{\circ}$ hemispherical model is aligned to the fisheye image in orthographic projection.}
\label{fig_ill_measure}
\end{figure}

\begin{figure*}
  \centering
  \begin{overpic}[width=\textwidth]{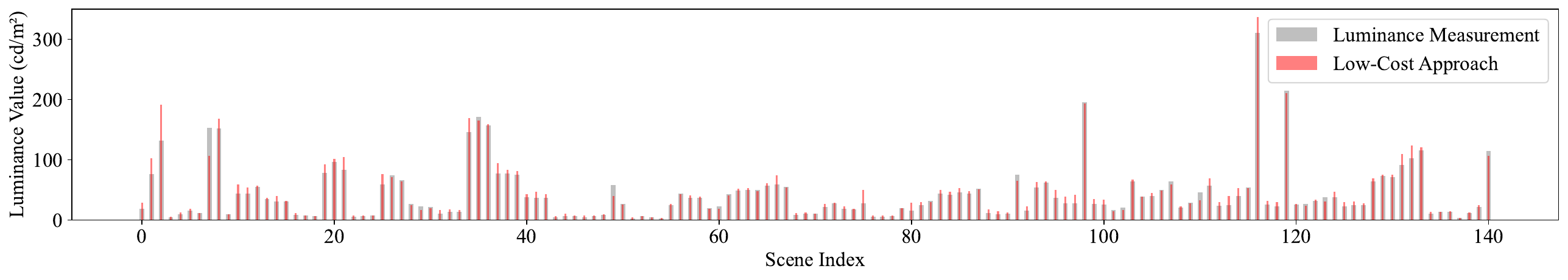}
  \end{overpic}
  \caption{Comparison of the luminance values on the whiteboard using the standard approach (Konica Minolta LS-160 luminance meter) and our low-cost approach (TS-710 light meter) for photometric calibration.}
  \label{fig_hdr_comp}
\end{figure*}

When capturing hemispherical HDR images, illuminance measurement has been used to calibrate fisheye HDR photographs~\cite {inanici2010evalution}. So far, illuminance calibration only applies to the fisheye image captured by 180$^{\circ}$ full-frame fisheye lens. The luminance meter measures radiance (luminance) emitted from a single point, while the light meter measures irradiance (illuminance) from 180$^{\circ}$ hemispherical directions. A single Konica Minolta LS-160 luminance meter costs over \$5{,}000\ US dollars, making photometric calibration expensive. In comparison, a TS-710 light meter (Fig.~\ref{fig_hdr_cali} (b)) costs around \$30\ US dollars. Our low-cost photometric calibration uses a light meter to find the calibration factor ($k$) and calibrate each HDR panorama. 

The Ricoh Theta Z1 camera captures two fisheye images, applies vignetting correction, and then merges them into an equirectangular image (Fig.~\ref{fig_ill_measure}(a)). To keep the camera pose consistent during data collection, we positioned a Ricoh Theta Z1 camera in the room, ensuring that one lens faced the main window opening while the lens with the digital display always faced backward. In this setup, for a single panorama with dimensions of $h$ (height) and $w$ (width), the pixel content within the range of $(0$:$h, \frac{w}{4}$:$\frac{3w}{4})$ is captured by the camera lens facing the front. In the meantime, as illustrated in Fig.~\ref{fig_hdr_cali}(b), we placed a TS-710 light meter next to the Ricoh Theta Z1 camera lens, facing the main window opening to measure the illuminance coming from 180$^{\circ}$ hemispherical directions. As the two camera lenses are not 180$^{\circ}$ hemispherical, the next step is to virtually convert the content captured by the front lens into a 180$^{\circ}$ orthographic projection.

A photograph captured by a fisheye lens has its own projection function and requires geometric transformation to correct the distortion caused by each lens~\cite{abreu2019method,bettonvil2005fisheye}. As shown in Fig.~\ref{fig_ill_measure}(b), the horizontal illuminance ($E$) is obtained by the luminance ($L$) from all 180$^{\circ}$ directions in the 3D world can be expressed as:
\begin{equation}\label{sphere_ill}
    \begin{split}
         {E} = {\int_0^{2\pi}}{\int_0^{\pi/2}}{L(\theta,\phi)}{\sin \theta}{\cos \theta}{d \theta}{d \phi}
    \end{split}
\end{equation}

where $L(\theta,\phi)$ represents the luminance from $(\theta,\phi)$. 

To get the uniform luminance ($L$), when {$L$ = $L(\theta,\phi)$}:
\begin{equation}\label{uni_ill}
    \begin{split}
         {L} = {E} / \pi
    \end{split}
\end{equation}

We crop the panorama and convert the content captured by the front lens into 180$^{\circ}$ orthographic projection. The measured illuminance value ($E$) is used to compute the uniform luminance ($L$). This uniform luminance will then be compared to the average luminance value from the orthographic HDR fisheye image to determine the calibration factor ($k$). Since the indoor and outdoor panoramas are captured simultaneously using the same camera and settings, the calibration factor ($k$) obtained from the indoor measurement can be used to calibrate the corresponding outdoor HDR image. The detailed process of panoramic HDR calibration is expressed in Alg.~\ref{alg:hdr_cali}. 

\begin{algorithm}
    \caption{Panoramic HDR Calibration}\label{alg:hdr_cali}
    \textbf{Input:} Captured HDR Panorama ($I$), Illuminace Value ($E$)
    \textbf{Output:} Calibrated HDR Panorama ($I'$)
    \begin{algorithmic}[1]   
        \State Crop the target region from $I$ and get an orthographic fisheye HDR image  
        \State Convert $E$ into uniform luminance value
        \State Compute luminance map for orthographic fisheye HDR image
        \State Obtain calibration factor between uniform luminance value and averaged displayed luminance value
        \State Apply calibration factor to $I$
    \end{algorithmic}
    
\end{algorithm}

\begin{figure}[!b]
  \centering
  \begin{overpic}[width=0.5\textwidth]{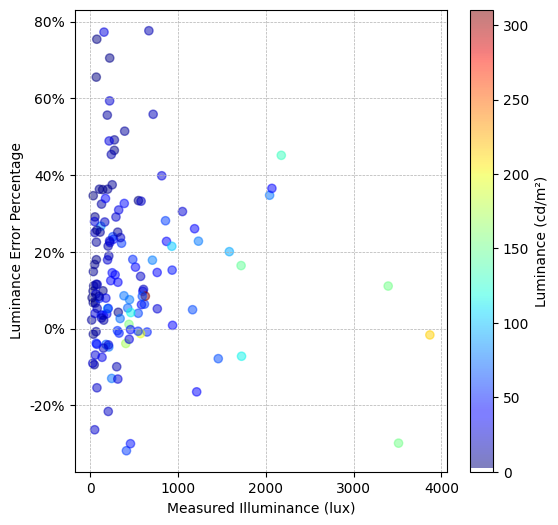}
  \end{overpic}
  \caption{Luminance error percentage on the whiteboard using the standard approach (Konica Minolta LS-160 luminance meter) and our low-cost approach (TS-710 light meter) for photometric calibration.}
  \label{fig_lum_perc_ill}
\end{figure}

During data collection, we used two approaches (Fig.~\ref{fig_hdr_cali}) to perform photometric calibration for each scene respectively. For the standard luminance measurement (Fig.~\ref{fig_hdr_cali}(a)), we placed a matte whiteboard as a target plane to measure the local luminance value. For comparison, we measured the illuminance value (Fig.~\ref{fig_hdr_cali}(b)) to calibrate the captured HDR panorama, then selected the target luminance value in the whiteboard region to compare with the value from standard luminance measurement. As a result (Fig.~\ref{fig_hdr_comp}), the absolute luminance error across all 141 scenes is 3.988 $cd/m^{2}$.

We plot the luminance error in percentage in Fig.~\ref{fig_lum_perc_ill}. The data points are color-coded based on luminance measured using the standard approach (Konica Minolta LS-160 luminance meter), shown on the right y-axis. The x-axis represents the measured illuminance from the low-cost approach (TS-710 light meter), while the left y-axis shows the error percentage after calibration with the low-cost approach. The plot suggests that when the measured illuminance is low, the percentage error tends to be higher. For instance, when the measured illuminance is 3 $lux$, but the actual value is 2 $lux$, this results in a 50\% error. Conversely, higher measured illuminance values generally correspond to lower percentage errors.

\section{Kitchen Remodeling Application}
\begin{figure}[!b] 
  \begin{overpic}[width=\linewidth]{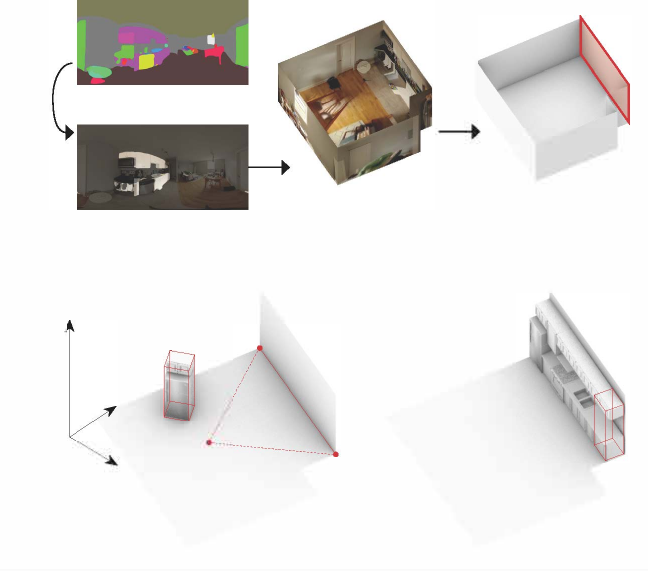}
    \put(1,70){\footnotesize \color{black}{(a)}}
    \put(12,72){\footnotesize \color{black}{Semantic Map}}
    \put(12,52.5){\footnotesize \color{black}{Kitchen Area}}
    \put(48,52.5){\footnotesize \color{black}{3D Layout}}
    \put(80,52.5){\footnotesize \color{black}{Kitchen Wall}}
    \put(20,0){\footnotesize \color{black}{Single Object}}
    \put(65,0){\footnotesize \color{black}{Multiple Objects}}
    \put(1,25){\footnotesize \color{black}{(b)}}
    \put(19,26){\footnotesize \color{black}{y}}
    \put(19,14){\footnotesize \color{black}{x}}
    \put(10,40){\footnotesize \color{black}{z}}
  \end{overpic}
   \caption{Kitchen Layout: (a) Using a single panorama as input, we first generate a semantic map to identify the kitchen area, and then the kitchen wall is segmented from this 3D layout. (b) The segmented kitchen wall and virtual objects are imported into the 3D coordinates. Multiple objects are then arranged along the kitchen wall based on geometric transformation.}
\label{fig_seg_kit}
\end{figure}
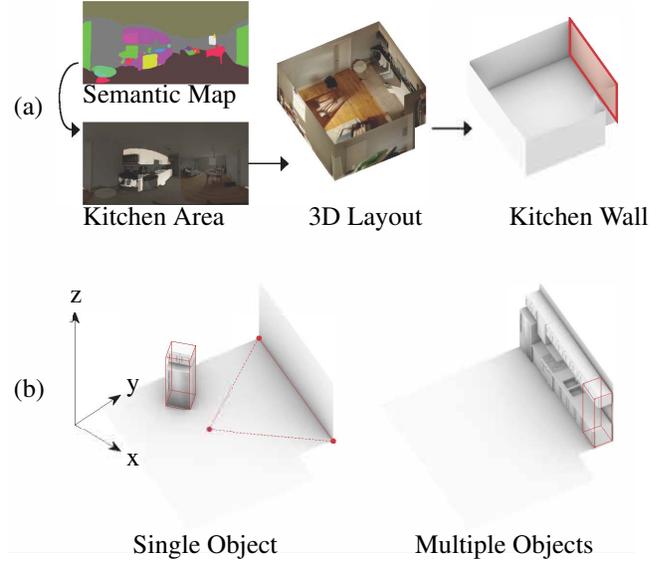

The kitchen remodeling process is based on scene geometry, where virtual kitchen components are inserted into the 3D floor layout for each scene. As illustrated in Fig.~\ref{fig_seg_kit}(a), we first use semantic segmentation~\cite{zhou2018semantic} to classify indoor objects into different categories and use the cabinet as an indicator to locate the kitchen area in the panorama. Meanwhile, the 3D floor layout is estimated from a single panorama~\cite{zhi2022semantically}, and the scene is segmented into individual planar surfaces. Finally, the kitchen area from the panorama is used to select the target walls for kitchen remodeling.

\subsection{Kitchen Layout Estimation}
Given the known 3D floor layout, each kitchen has a unique dimension and configuration. We first modeled the detailed geometry for the cabinets and appliances to insert virtual objects for kitchen remodeling. In 3D modeling, each object is assembled from individual components, such as handles, cabinets, and countertops, allowing for further material customization and scene editing.

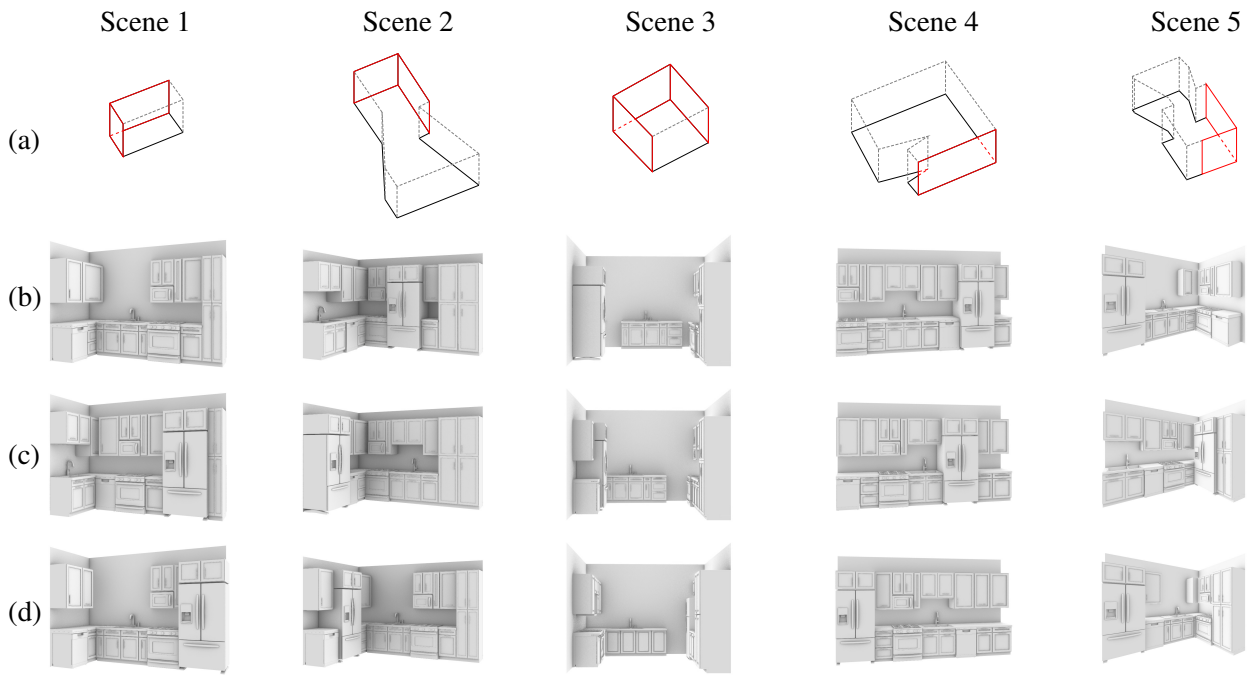
\begin{figure*} 
  \begin{overpic}[width=\linewidth]{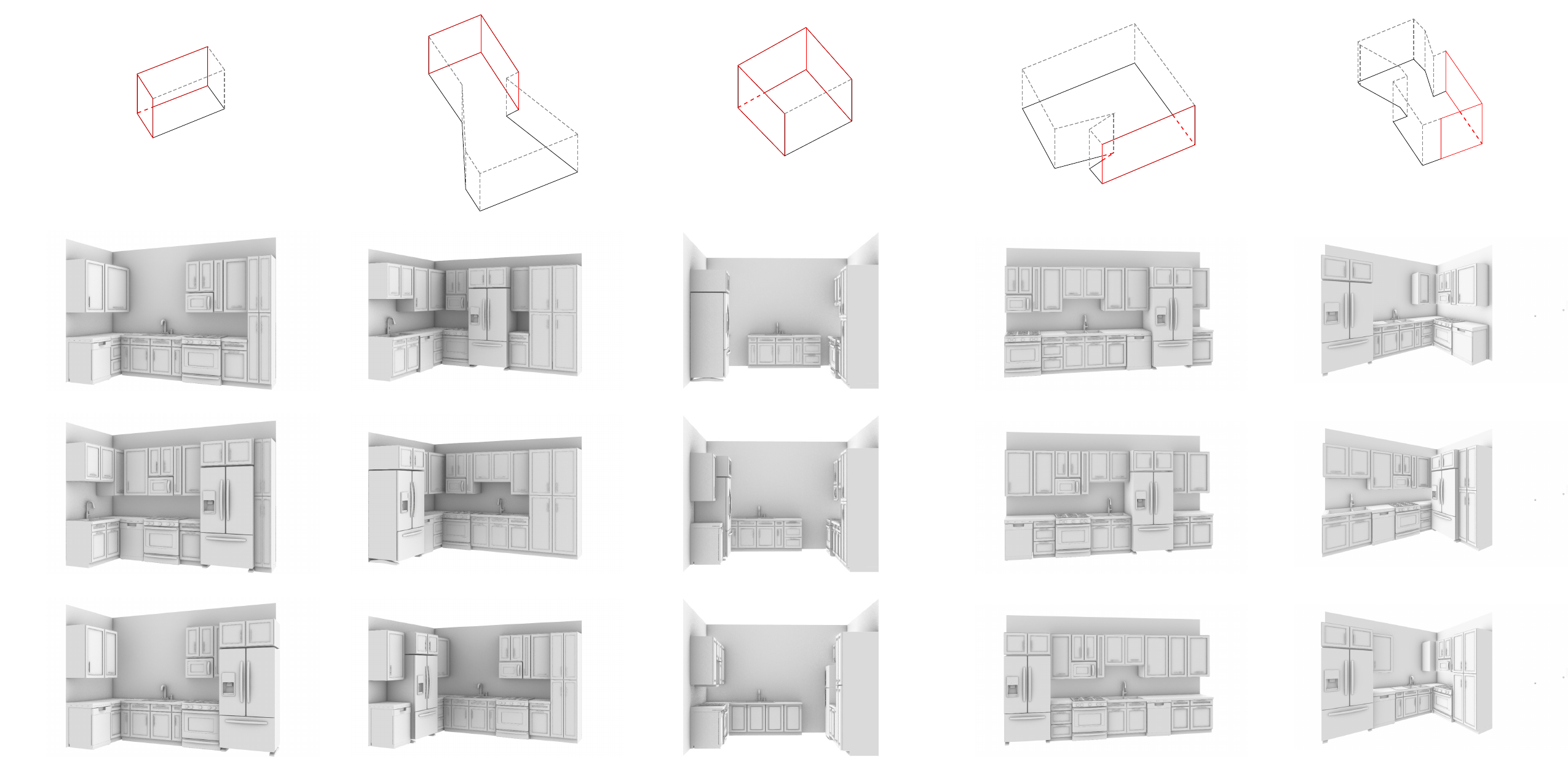}
    \put(1,41){\small \color{black}{(a)}}
    \put(1,29){\small \color{black}{(b)}}
    \put(1,17){\small \color{black}{(c)}}
    \put(1,5){\small \color{black}{(d)}}
    \put(8,50){\small \color{black}{Scene 1}}
    \put(28,50){\small \color{black}{Scene 2}}
    \put(48,50){\small \color{black}{Scene 3}}
    \put(68,50){\small \color{black}{Scene 4}}
    \put(88,50){\small \color{black}{Scene 5}}
  \end{overpic}
   \caption{Object Insertion and Kitchen Layouts: (a) The kitchen walls, highlighted in \textcolor{red}{Red}, are segmented from the estimated 3D layout as planar surfaces. Virtual kitchen components are inserted into the kitchen space with various layout alternatives, as shown in (b), (c), and (d).}
\label{fig_layout_kit}
\end{figure*}

To insert a single new object (Fig.~\ref{fig_seg_kit}(b)), the kitchen wall and floor mesh are imported into 3D coordinates, where the kitchen wall corners and the centroid point of the floor mesh are located. The new object is imported with a bounding box to determine its spatial dimension and orientation. The object will be rotated around the $z$-axis by angle $\theta$ to adjust its orientation to align with the target kitchen wall, then translated by $t_x$ in the $x$-axis and $t_y$ in the $y$-axis. The 4 by 4 transformation matrix ($M$) is defined by $M = R_z(\theta) \cdot T(t_x, t_y)$, and it is applied to the original object with the 3D point set $\mathbf{x_i}$. This transformation maps each point to its corresponding transformed point $\mathbf{x_i}'$ in the $xy$ plane. 



Given multiple objects, each object is arranged in a sequence order defined by user inputs, like $Refrigerator$, $Cabinet$, $Oven$, $Sink$, $Cabinet$ in Fig.~\ref{fig_seg_kit}(b). Each object will be placed side by side along the kitchen wall. Since the actual wall length is unknown, the last object needs to be slightly scaled in width to cover the wall corner; the wall corner can also be left unfurnished. The detailed kitchen layout process is expressed in Alg.~\ref{alg:kit_place}.

\begin{algorithm}
    \caption{Kitchen Layout}\label{alg:kit_place}
    \textbf{Input:} 3D Layout, Kitchen Components, Sequence Order
    \textbf{Output:} Transformed Kitchen Components
    \begin{algorithmic}[1]   
        \State identify the layout type of kitchen space
        \For{$i$ in Kitchen Components}
            \State get 3D dimension and adjust the orientation 
            \State find the target location from Sequence Order
            \State compute and apply the transformation matrix
        \EndFor
    \end{algorithmic}
\end{algorithm}

The kitchen spaces exhibit different layout typologies in each scene. Figure \ref{fig_layout_kit} showcases examples of kitchen layouts and alternatives for object insertions. Based on the identified kitchen spaces, the layout types are classified as follows: a single wall forms an $I$ shape, two continuous walls with one corner form an $L$ shape, and three sides with two corners create a $U$ shape. For layout design, the $L$ shape and $U$ shape can be managed by dividing the walls into multiple linear sections, allowing kitchen components to be placed linearly along each wall. Our algorithm adapts to various kitchen layout alternatives with different orientations and arranges virtual kitchen components along the walls in different sequences. Additionally, the layout design can accommodate individual cases, leaving empty spaces in the corner areas.

\subsection{Indoor Virtual Staging}
\begin{figure*}
  \centering
  \begin{overpic}[width=0.95\textwidth]{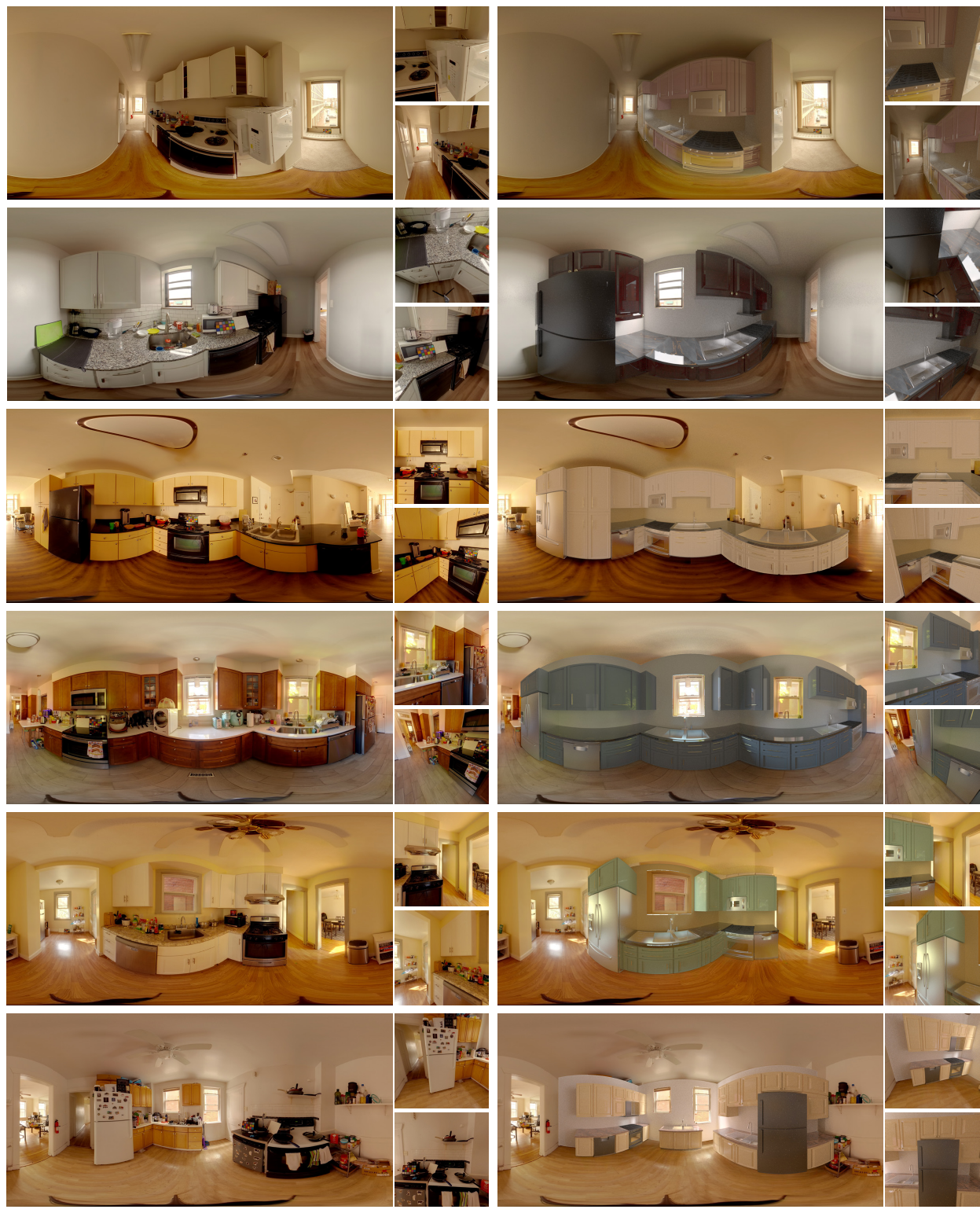}
    \put(1,90){\small{\makebox(0,0){\rotatebox{90}{Scene 1}}}}
    \put(1,73.5){\small{\makebox(0,0){\rotatebox{90}{Scene 2}}}}
    \put(1,57.5){\small{\makebox(0,0){\rotatebox{90}{Scene 3}}}}
    \put(1,41){\small{\makebox(0,0){\rotatebox{90}{Scene 4}}}}
    \put(1,25){\small{\makebox(0,0){\rotatebox{90}{Scene 5}}}}   
    \put(1,9){\small{\makebox(0,0){\rotatebox{90}{Scene 6}}}}
    \put(13,98.5){\small \color{black}{Captured Scenes}}
    \put(55,98.5){\small \color{black}{Edited Scenes}}
  \end{overpic}
  \caption{Photo Gallery of Kitchen Remodeling: (left) The captured scenes showcase various existing kitchen spaces. (right) Using our application, the kitchen walls are repainted with new colors, and the kitchen spaces are remodeled with new appliances, cabinets, and countertops. The newly edited kitchens are virtually inserted into the scenes.}
  \label{fig_vir_staging}
\end{figure*}

We developed an editable rendering pipeline using the Mitsuba rendering system~\cite{jakob2022mitsuba3} to achieve high-quality virtual staging. Using paired indoor-outdoor panoramas, virtual staging can relight the existing scene with new kitchen components with real-time spatially-varying outdoor light and global illumination (Fig.~\ref{fig_vir_staging}). For each scene, the rendering pipeline automatically takes the 3D layout, scene textures, and material properties, along with the outdoor panoramas, as inputs to generate a new virtual panorama. 

Individual kitchen components, such as cabinets, countertops, and handles, are paired with corresponding material properties and textures to generate an accurate appearance as real-world materials. By incorporating the complete 3D scene geometry and 360$^{\circ}$ environment map, the rendering model can capture light from multiple directions, enabling the detailed outdoor scene to be reflected on indoor surfaces such as cabinets. 

\begin{figure*}
  \begin{overpic}[width=\linewidth]{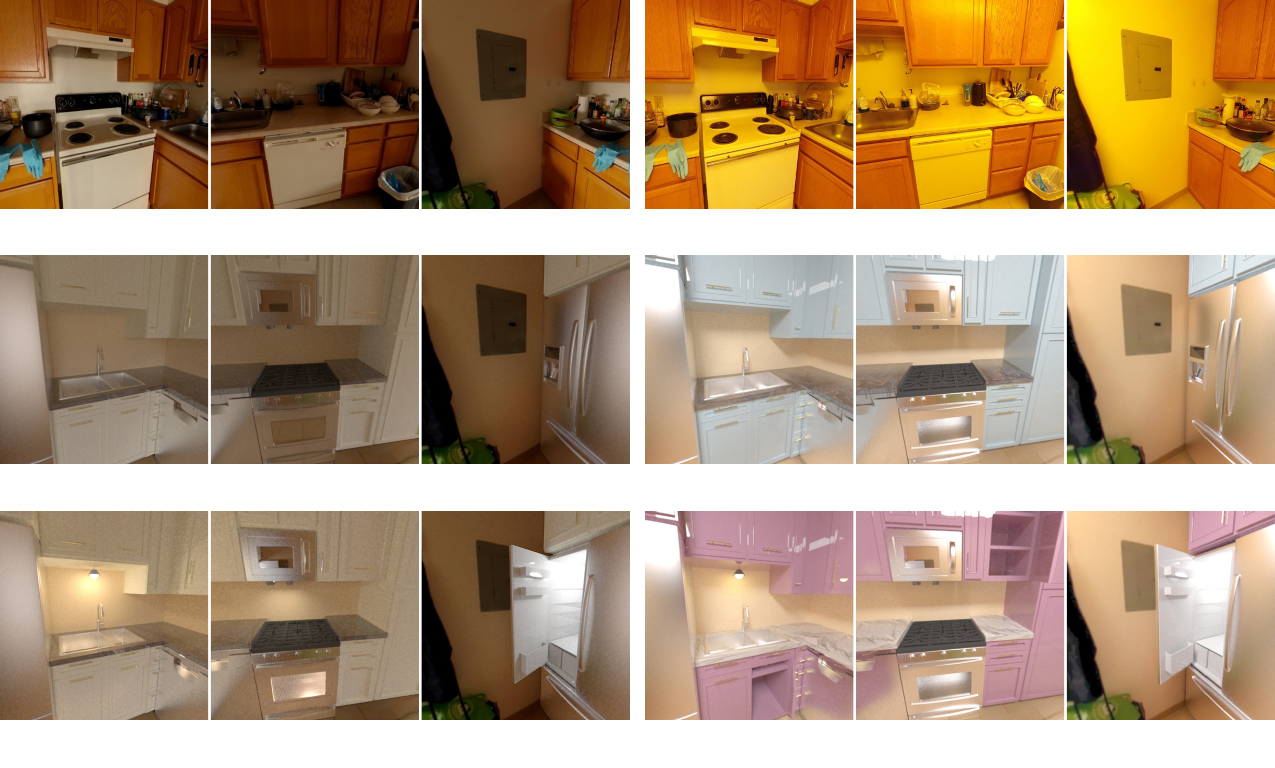}
      \put(23,42){\small \color{black}{(a)}}
      \put(23,22){\small \color{black}{(b)}}
      \put(23,2){\small \color{black}{(c)}}
      \put(73,42){\small \color{black}{(d)}}
      \put(73,22){\small \color{black}{(e)}}
      \put(73,2){\small \color{black}{(f)}}    
  \end{overpic}
  \caption{Editing the Existing Scene with New Kitchen Components and Electrical Lighting: (a) Scene captured under natural illumination. (b) The existing scene remodeled with new kitchen components. (c) Addition of electrical lighting in the kitchen space. (d) The existing scene captured under natural illumination and the existing electrical lighting as a reference. (e) Remodeled kitchen space illuminated by additional ceiling light with accurate spectrum (Endura OT16-3101-WT MR16: LED - 6336K - 76.00 CRI)~\cite{SPD_curve}. (f) The scene rendered with new cabinets and countertops.}
\label{fig_kit_material}
\end{figure*}

Compared to the previous data-driven approaches~\cite{wu2023factorized,li2022physically,gardner2017learning,dastjerdi2023everlight}, our rendering pipeline provides flexibility in editing scene materials, scene geometry, and lighting conditions for complex scenes. In Fig.~\ref{fig_kit_material} (a-c), the captured panorama showcases the existing kitchen space under natural illumination, which is then remodeled with new kitchen components and lighting fixtures. In Fig.\ref{fig_kit_material} (d-f), the same scene was captured with electrical lighting during the daytime as a reference. Subsequently, ceiling light with an accurate spectrum (Endura OT16-3101-WT MR16: LED - 6336K - 76.00 CRI)\cite{SPD_curve} is added to illuminate the new kitchen space, and the materials for the cabinets and countertops are modified to render a new scene.

\section{Discussion}
\subsection{Summary}
This paper introduces a virtual staging application that captures existing scenes, edits kitchen spaces, and relights them with virtual objects featuring customized materials. Furthermore, we contribute a Pano-Pano HDR dataset with a photometric calibration method and a detailed kitchen component dataset for future research. Specifically, we provide two applications:

\textbf{Panoramic HDR Calibration} We proposed an HDR photography approach for collecting paired indoor-outdoor panoramas using a single camera. Our low-cost calibration method enables the photometric calibration of HDR photographs with absolute photometric measurements. Compared to the standard approach, our method offers an affordable solution for recovering scene radiance in different indoor and outdoor scenarios, making it applicable to a larger user base in the future.

\textbf{Indoor Virtual Remodeling} We developed an inverse rendering pipeline for editing kitchen spaces and remodeling the existing scenes with new kitchen layouts. The application constructs a full rendering model from captured photographs and allows for high-quality virtual staging through physics-based rendering. New virtual objects are rendered within the scene geometry with real-time spatially varying light. The application offers editable materials, indoor layouts, and light estimation for the existing scenes. The outcome provides a convenient tool for the general public and housing agents in the real estate market.

\subsection{Limitations and Future Work} This study has several limitations. Based on our dataset, kitchen layouts are classified into three main shapes: $I$, $L$, and $U$. Future work needs to evaluate other kitchen layouts to ensure comprehensive coverage of different scenarios. Besides, while we modeled new kitchen components for virtual remodeling, future efforts should expand this dataset to include more kitchen components and materials. Extensive usability testing with end-users is necessary to gather feedback on the ease of use and functionality of our applications. These aspects will be addressed in future work.

{\small
\bibliographystyle{ieee_fullname}
\bibliography{egbib}
}

\end{document}